\documentclass[doublecol]{epl2} % for 2 columns style with line numbers
\usepackage{graphicx}
\usepackage{bm}
\usepackage{pdfpages}
\input{epsf}
\usepackage{color}

\renewcommand{\emph}[1]{{\it #1}}

\newcommand{\beq}{\begin{equation}}
\newcommand{\eeq}{\end{equation}}
\newcommand{\beqa}{\begin{eqnarray}}
\newcommand{\eeqa}{\end{eqnarray}}

\title {Local rectification of heat flux}

\author{M. Pons\inst{1} \and Y. Y. Cui\inst{2} \and A. Ruschhaupt\inst{3} \and M. A. Sim\'on\inst{2} \and J. G. Muga\inst{2}}
\shortauthor{M. Pons \etal}

\institute{
  \inst{1} Departmento de F\' isica Aplicada I, Universidad del Pa\' is Vasco, UPV-EHU, Bilbao, Spain\\
  \inst{2} Departmento de Qu\' imica-F\' isica, Universidad del Pa\' is Vasco, UPV-EHU, Bilbao, Spain\\
  \inst{3} Department of Physics, University College Cork, Cork, Ireland
}

\abstract{We present a chain-of-atoms model where heat is rectified, with different fluxes from the hot to the cold baths
located at the chain boundaries when the temperature bias is reversed. The chain is homogeneous except for boundary effects
and a local modification of the interactions
at one site, the ``impurity''. The rectification mechanism is due here to the localized impurity, the only asymmetrical element of the structure, apart from the externally imposed temperature bias,  and does not rely on putting in contact different materials or other known mechanisms such as grading or long-range interactions.  The effect survives if all interaction forces are linear except the ones for the impurity.}

\begin{document}

\maketitle

\section{Introduction}

In spite of much work on thermal rectification
after the first model proposal in 2002 \cite{Casati2002} (for a broad perspective on heat rectification
see \cite{Roberts2011,Li2012}), the manipulation of phononic
heat fluxes is still far from being completely controlled as no efficient and feasible thermal diodes have been found \cite{Casati2015,Pereira2017}.
The thermal rectifier, a device where the heat
current changes when the temperature bias of the thermal baths at the boundaries is reversed,
is one of the key tools needed to manipulate heat currents and build thermal circuits.
A wealth of research is underway to meet the challenge posed by  a ``near standstill'' of the field \cite{Casati2015,Pereira2017},
combined with the prospects of widespread and impactful practical applications.
Together with experimental progress, at this stage work exploring new models is important to test possibilities that may become feasible as control capabilities improve \cite{Roberts2011}.
In this paper we propose,
motivated by previous work on ``atom diodes"  \cite{Muga2004,Raizen}, a rectifying scheme
based on the effect of a local defect, or impurity,
in an otherwise homogeneous system.

To date, there have been several proposals of systems that could be used to rectify heat flows at the nanoscale.
A common scheme is based on coupling two or more different homogeneous segments, modelled with
chains of atoms with nonlinear interactions (which in this context means non-linear forces, i.e., anharmonic potentials) \cite{Casati2002,Casati2004,Hu2006,Peyrard2006,Benenti2016}
or with a temperature and position dependent
conductivity assuming expressions for the heat current \cite{Peyrard2006,HHYZ}.
%Other proposals \cite{Hu2006} make use of a two-segment Frenkel Kontorova model \cite {Casati2004} finding size effects such us the reversal of the rectification as the size of the system increases, depending on the strength of the coupling between both segments.
Basic ingredients for thermal rectification have been considered to be  the asymmetry in the system and non-linear interactions
\cite{Zeng2008,Li2012,Benenti2016}, which lead to a temperature dependence of the phonon bands, or power spectral densities,  of the
weakly coupled \cite{Hu2006} segments. These bands match or mismatch at the interfaces, depending of the sign of the temperature bias, leading, respectively, to heat flow or insulating conditions. In fact, alternative mechanisms due to band mixing appear for stronger coupling
or long chains \cite{Hu2006}, and Pereira \cite{Pereira2017},  based on minimalist models,
has recently reformulated the conditions leading to rectification  as the combination of asymmetry plus the existence of some feature of the system that depends on the temperature (nonlinearity is certainly a possible cause of such dependence).
Other systems proposed are graded materials, such as a chain with an uneven distribution of mass \cite{Chang2006,Zeng2008,Casati2015}, and long-range interactions  have been shown to be able to amplify the rectification and avoid or mitigate the decay of the effect with system size \cite{Pereira2013,Casati2015}.
%Mass graded materials can be built, but up to now the rectification experimentally observed is small \cite{Chang2006}.
Also, recent models and experiments use asymmetrical homogeneous or inhomogeneous nanostructures and, in particular, graphene  \cite{nano,Wang2017}.
Finally, we mention for completeness theoretical works farther from our model
that consider the use of a quantum system \cite{Roberts2011}, such a three-level system, with each level coupled to a thermal bath \cite{Joulain2016}, or a double well with different frequencies to implement the asymmetry \cite{Koslov2016}.

The model proposed in this paper is a one-dimensional chain of atoms where all, except one of them, are trapped in on-site harmonic potentials and interact with their nearest neighbours by Morse potentials (or also by harmonic
potentials in a simplified version). Unlike most chain models, the structural asymmetry is only in the impurity, which is subjected to a different on-site potential and interaction with one of its neighbors. The chain is connected to thermal baths at different temperatures at the boundaries.
%each end, and we have chosen the well known Nos\'e-Hoover model to simulate the thermal baths \cite{Martyna1968}.
%The manipulation of the potentials that affect one of the atoms makes the excitations easier to be transferred from one side, providing a mechanism for the inhomogeneity needed.

First, we shall describe the
%initial system used, that consists of a
homogeneous 1D chain, without the impurity,  %For this system, we numerically solve the dynamical equations, to show that the usual heat conduction applies. Once the system is connected to the baths at different temperature, after a transient time, the steady regime with constant value of the heat flux, $J$, is reached, and
and show that the Fourier law is fulfilled.
%Our goal is to obtain a system that presents a direction dependent heat flux when the positions of the baths are interchanged. With this aim,
Then  we modify the potentials for one of the atoms and demonstrate the rectification effect.

\section{Homogeneous one-dimensional chain}

We start with a homogeneous $1D$ chain with $N$ atoms coupled at both extremes to heat baths, at different temperatures $T_h$ and $T_c$ for ``hot" and ``cold" respectively. The baths are modeled with a Nos\' e-Hoover method as described in \cite{Martyna1968}.
Atoms $1$ and $N$ represent the first and the $N$-th atom in the chain, from left to right,
that will be in contact with the baths. All the atoms are subjected to on-site potentials and to nearest-neighbor interactions, and their equilibrium positions $y_{i0}$ are assumed to be equally spaced by a distance $a$.  $x_i= y_i-y_{i0}$,
$i=1,...,N$, represent the displacements from the equilibrium positions of the corresponding atoms
with positions $y_i$.

The classical Hamiltonian of the atom chain can be written in a general form as
\begin{equation}
\label{GH}
%GH=general Hamiltonian
H=\sum_{i=1}^{N} H_i,
\end{equation}
with
\begin{eqnarray}
\label{GH2}
%GH=general Hamiltonian
H_1&=&{{p^2_1} \over {2m}} +U_1(x_1)+V_L,
\nonumber\\
H_i&=&{{p^2_i} \over {2m}} +U_i(x_i)+V_i(x_{i-1},x_i)  \quad i=2,...,N-1,
 \nonumber\\
H_N&=&{{p^2_N} \over {2m}} +U_N(x_N)+V_N(x_{N-1},x_N) + V_R,
\end{eqnarray}
where the $p_i$ are the momenta, $U_i(x_i)$ is the on-site potential for the $i$th atom, and $V_i(x_{i-1},x_i)$ represents the atom-atom interaction potential. $V_R$ and $V_L$ are the interactions coupling the boundary atoms to the Nos\'e-Hoover thermostats, see \cite{Martyna1968}.

There are a large number of $1D$ models that obey this general Hamiltonian. Using different potentials for the trapping and interactions we would get different conductivity behaviors.
%%%%%%%%%%%%%%%%%%%%%%%
\begin{figure}
\centering
\includegraphics[width=8.8cm]{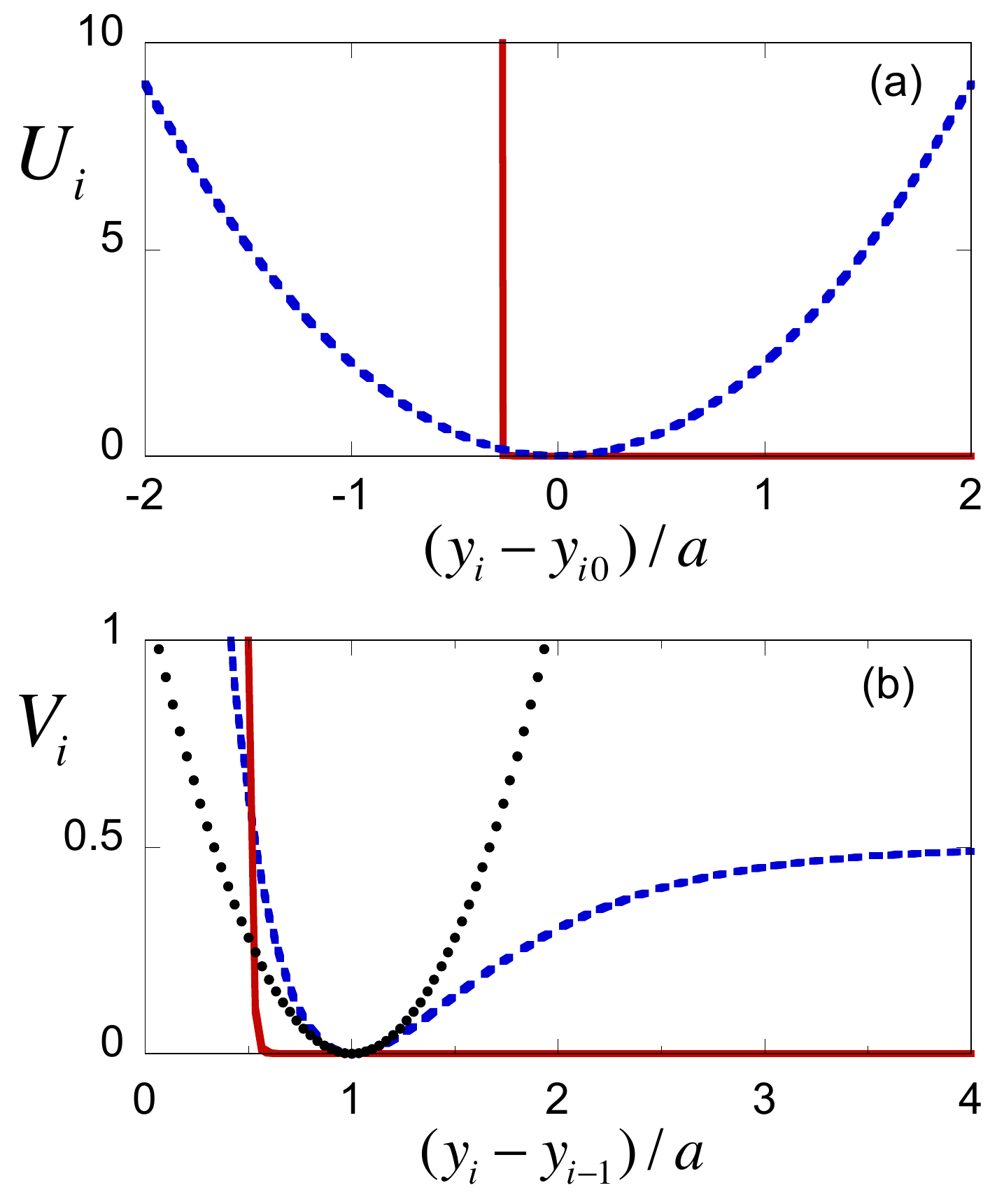}
\caption{(Color online) (a) On-site potentials: harmonic potential centered at the equilibrium position of each atom (dashed blue line) as a function of the displacement from this position $x_i=y_i-y_{i0}$ in $a-$units, and the on-site potential for the impurity, $i=N/2+1$
($N$ even, red solid line). (b) Interaction potentials as a function of the distance between nearest neighbors: Morse potential
(blue dashed line) valid for all atoms except for $i=N/2+1$, $N$ even, where the modified potential (red solid line) is used.
The harmonic approximation of the Morse potential is also depicted (eq. (\ref{Vhar}), black dots, only used for fig. 5, below).
Parameters: $D=0.5$, $g=1$, $\gamma = 45$, $d=100$ and $b=105$, used throughout the paper.
}
\label{figure1}
\end{figure}
%%%%%%%%%%%%%%%%%%%%%%%%%
%
We choose a simple form of the Hamiltonian in
%(\ref{IH})
which each atom is subjected to a harmonic on-site potential and a Morse interaction potential between nearest neighbors (see fig. \ref{figure1}, dashed lines),
\beqa
\label{HO}
%HO=Harmonic oscillator
U_i(x_i)&=&{1 \over 2} m \omega^2 x^2_i,
%\end{equation}
%\begin{equation}
\\
\label{IH}
%IP=Interaction potential
V_i(x_{i-1},x_i)&=&D\left \{e^{-\alpha [x_i-x_{i-1}]}-1\right \}^2,
\eeqa
where $\omega$ is the trapping angular frequency, and $D$ and $\alpha$ are time independent parameters of the Morse potential.
A ``minimalist version'' of the model where $V$ becomes the harmonic limit of eq. (\ref{IH}), dotted line in fig. 1,
 will also be considered in the final discussion,
\beq
\label{Vhar}
{V}_i(x_{i-1},x_i)=k(x_i-x_{i-1})^2/2,\;k=2D\alpha^2.
\eeq
For convenience, dimensionless units are used and the mass of all particles is set to unity.
%%%%%%%%%%%%%%%%%%%%%%%%%%%%
\begin{figure}
\centering
\includegraphics[width=8.8cm]{{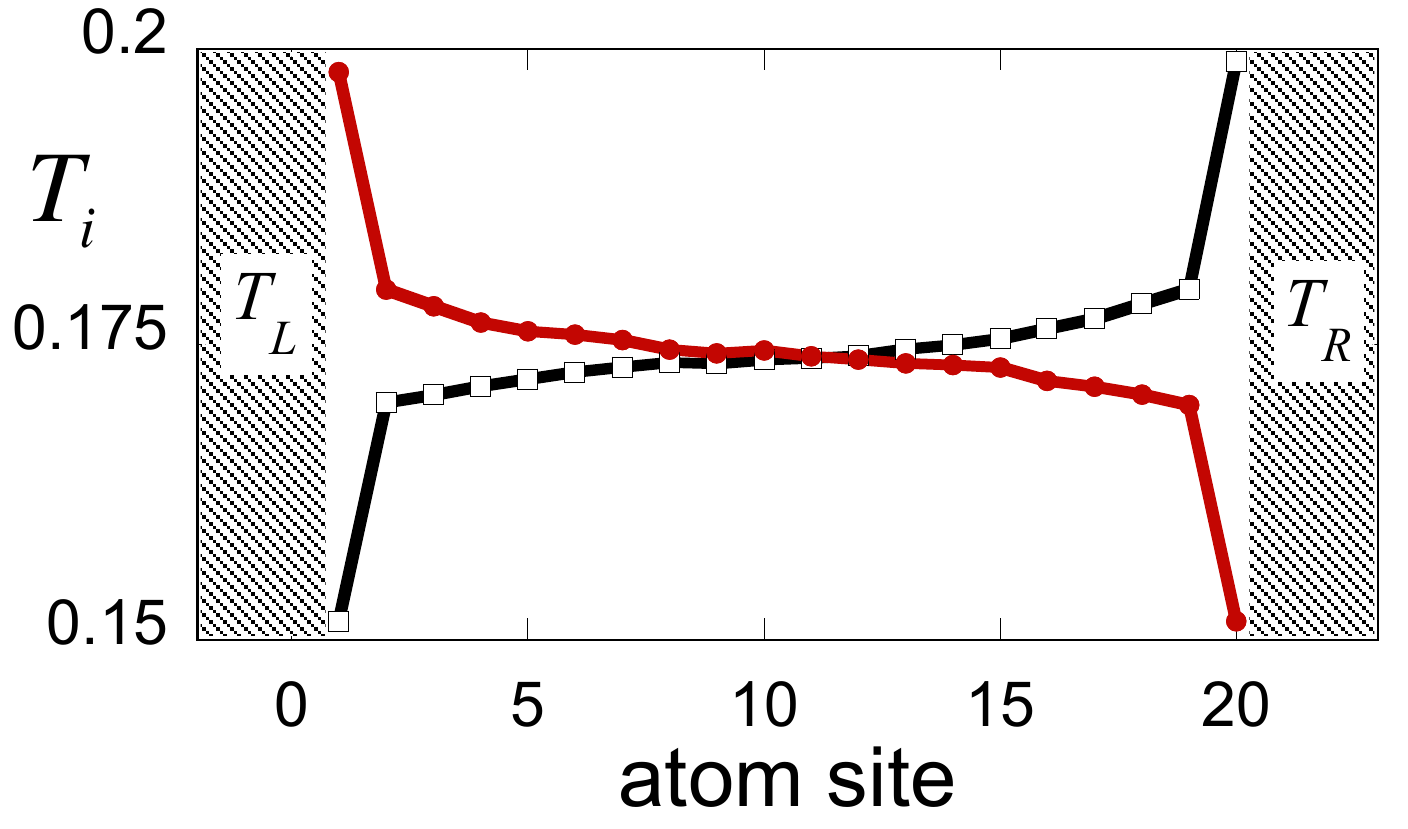}}
\caption{(Color online) Symmetric temperature profiles for a homogeneous chain, without impurity.  For $T_{h}=T_{L}$, $T_c=T_R$ (red solid dots) the (absolute value of) the heat flux is $J_{L\rightarrow R}$, equal to $J_{R\rightarrow L}$ for the reverse configuration of the bath temperatures, $T_{h}=T_{R}$, $T_c=T_L$
(black empty squares). Parameters as in fig. 1.}
\label{figure2}
\end{figure}
%%%%%%%%%%%%%%%%%%%%%%%%%%%%%%

We start by studying the homogeneous configuration with no impurity and potentials (\ref{HO}) and (\ref{IH}), solving numerically the dynamical equations for  the Hamiltonian (\ref{GH}) with a Runge-Kutta-Fehlberg algorithm. We have chosen a low number of atoms, $N=20$,  with thermal baths at $T_h=0.20$ and $T_c=0.15$ at both ends of the chain with 16 thermostats each. The real temperature is related to the dimensionless one through $T_{real}=T m a^2 \omega^2/k_B$ so, for typical values  $m\approx10^{-26}$ kg, $\omega \approx 10^{13}$ s$^{-1}$, $a\approx 10^{-10}$ m, and using $k_B =1.38 \times 10^{-23}$ JK$^{-1}$,
the dimensionless temperatures $0.15,\, 0.20$, translate into $100,\, 150$ K. It is advisable to use temperatures around these values so that we ensure that the displacements of the particles are realistic \cite{Casati1984}.

First we demonstrate  that our system satisfies Fourier's heat law for the heat flux, $J=\kappa \nabla T$.
%, so it shows normal thermal conductivity. ESTE CONCEPTO TIENE QUE VER CON EL TAMA�O?
To this end, we calculate the local heat flux $J_i$ and temperature $T_i$, performing the numerical integration
%of Eq. (\ref{GH2})
for long enough times to reach the stationary state.
The local temperature is found as the time average $T_i= \langle p_i^2 / m \rangle$, whereas
%After a transient, the local temperature is given by the time average $T_i=\langle p_i^2\rangle$.
$J_i$,  from the continuity equation
%, $\dot H(x,t)+divJ(x,t)=0,$
\cite{Hu1998}, is given by
%
%Fourier law: temperature gradient vanishes with N
\begin{equation}
\label{heatflux}
J_i=-\dot x_i {{\partial V(x_{i-1},x_{i})} \over {\partial x_i}}.
\end{equation}
From now on we only consider the time average $\langle J_i (t)\rangle$, which converges to a constant value for all sites once the system is in the stationary nonequilibrium state. We depict the temperature profiles, for $N=20$, first with $T_L=T_h$ and $T_R=T_c$
($L$ and $R$ stand for left and right) and after switching the positions of the thermal baths in fig. \ref{figure2}. The profiles are symmetric, as expected, and the heat flux does not have a preferred direction  \cite{Hu1998,Casati2002}. Denoting the absolute values of the fluxes from the left (when $T_L=T_h$) as
$J_{L\rightarrow R}$, and from the right (when $T_R=T_h$) as
$J_{R\rightarrow L}$, we find that $J_{L\rightarrow R}=J_{R\rightarrow L}=J=1.6\times 10^{-2}$, in the dimensionless units, consistent with the values found in other models \cite{Casati2002,Hu1998}.
%%%%%%%%%%%%%%%%%%%%%%%%%%%%%%%%%%
\begin{figure}
\centering
\includegraphics[width=8.8cm]{{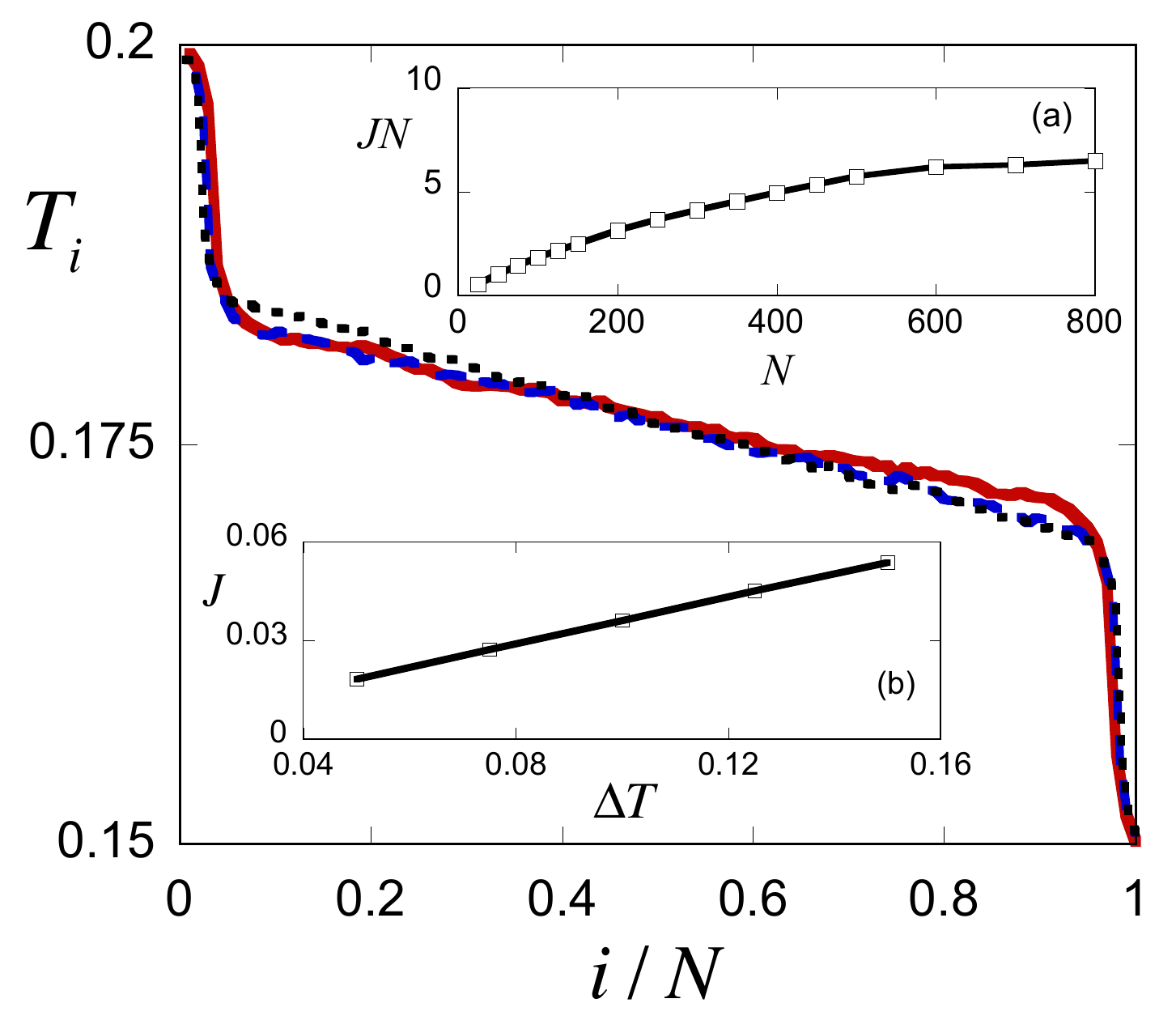}}
\caption{(Color online) Temperature profile along the homogeneous chain for different number of atoms: 100 (dotted black line), 125 (dashed blue line) and 150 (solid red line). The atom sites have been rescaled with the total number of atoms, showing the convergence of the spatial profile of the local temperature $T_i$. The time averages have been carried over a time interval of $\approx 2 \times 10^6$ after a transient of $\approx 1\times 10^5$. In the inset (a), the product $JN$ vs. $N$ demonstrates that for long chains $JN$ is independent of $N$. In (b) the linear dependence of $J$ with $\Delta T$ for a fixed number of atoms, $N=100$, is shown. Parameters as in fig. 1.}
\label{figure3}
\end{figure}
%%%%%%%%%%%%%%%%%%%%%%%%%%%%%%%%%%%%%%%%

The profile of the temperature is linear with nonlinearities at the edges, close to the thermal baths,  due to the boundary conditions \cite{Lepri1997}. In fig. \ref{figure3} we depict $T_i$ vs $i/N$ for $N=100, 125$ and $150$ with the same boundary conditions. For these
larger atom numbers  we have connected the first 3 and the last 3 atoms to the Nos\'e-Hoover baths. The temperature gradient scales as $N^{-1}$, which is also true for many other different models \cite{Hu1998}. In the inset (a) of fig. \ref{figure3}  the product $JN$ is plotted vs. $N$ showing that in a low $N$ limit there is a well defined conductivity per unit length whereas for longer chains, $JN$ tends to be constant  which indicates a normal thermal conductivity independent of the length. Fixing the number of atoms to 100, as in the inset (b) of fig. \ref{figure3},  we observe that the Fourier law, $J=\kappa \nabla T$, is fulfilled.

\section{Impurity-based thermal rectifier}

To rectify the heat flux
we modify the potentials for site $j=N/2+1$ with $N$ even, as
%
%\begin{equation}
%\label{IMP1}
%IMP1=impurity in absolute position
%U_j(y_j,t)=d e^{-b [y_j(t)-y_{d}]} +ge^{-\gamma [y_j(t)-y_{j-1}(t)-\epsilon]}
%\end{equation}
%with $y_{d}=y_{d,0}-a/3$.  Written in terms of the displacements, $x_j$,
%
\beqa
\label{IMP}
%IMP=impurity
U_j(x_j,t)&=&d e^{-b [x_j(t)+a/3]},
\\
V_j(x_{j-1},x_j,t)&=&ge^{-\gamma [x_j(t)-x_{j-1}(t)+a/2]}.
\eeqa
All the parameters involved, $d, b$, and $g,\gamma$ are time independent. In fig. \ref{figure1} the modifications introduced with respect to the ordinary sites are shown (solid lines).  The different on-site and interaction terms introduce soft-wall potentials
(instead of hard-walls to aid integrating the dynamical equations) that make it difficult for the impurity to transmit its excitation to the left whereas left-to-right transmission is still possible.
This effect is facilitated by the relative size of the coefficients, $a/3<a/2$, that determine the position of the walls.
% that we fixed after some experimentation.
They imply that an impurity excited by a hot right bath cannot affect its left cold neighbour near its equilibrium position at the $j-1$ site.
However, if the left $j-1$ atom is excited from a hot bath on the left,
it can get close enough to the impurity to kick it and transfer kinetic energy.
The asymmetrical behavior relies on the asymmetry of the potentials and the temperatures of
neighboring atoms; it does not require
breaking time-reversal invariance. All collisions are elastic and time-reversible.

After extensive numerical simulations, we have chosen the values of these parameters as in fig. 1, such that the conductivity in the forward direction, $J_{L\rightarrow R}$, and the rectification factor, defined as $R=(J_{L\rightarrow R}-J_{R\rightarrow L}) / J_{R\rightarrow L}\times 100$,
are both large for $T_h=0.2$, $T_c=0.15$. A large $R$ without a large $J_{L\rightarrow R}$ could in fact be useless \cite{Roberts2011}.
%($R=0$ would represent a perfectly symmetric heat conduction.).
Note that this is not necessarily the optimal combination, which in any case would depend on the exact definition of ``optimal'' (technically on how $J_{L\rightarrow R}/J$ and $R$ are weighted and combined in a cost function and on the limits imposed on the
parameter values). This is an interesting problem but it goes beyond the focus of our paper,
which is to demonstrate and discuss the effect.

\begin{figure}
\centering
\includegraphics[width=8.8cm]{{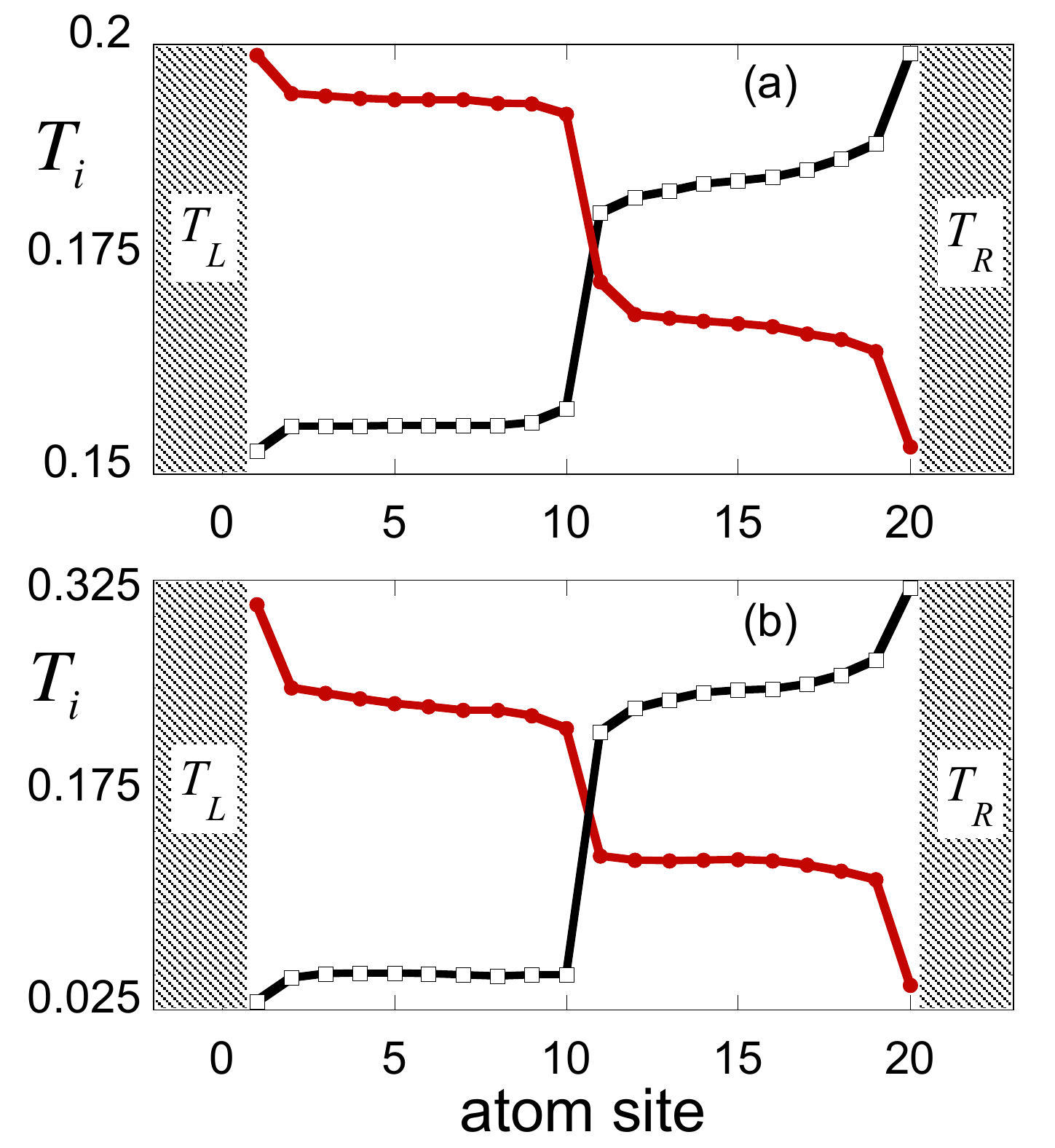}}
%(b)\includegraphics[width=8.8cm]{{FIG6.pdf}}
%\includegraphics[width=8.8cm,height=6cm]{{FIG1b.pdf}}
\caption{(Color online) Temperature profile for the chain of $N=20$ atoms, with an impurity in the $N/2+1$ position, with $T_L=T_h$ and $T_R=T_c$ (circles) and with the thermostat baths switched (squares).
%, for $g=1, \gamma = 45, d=100$ and $b=105.$ The rest of parameters are the same as in fig. 3 where the profile of the temperature with no impurity was shown. The difference in the temperature profiles can be clearly noticed, also confirmed by the heat fluxes
Parameters as in fig. 1.
(a) $T_c=0.15$, $T_h=0.2$. $J_{L\rightarrow R}=0.00769$ vs $J_{R\rightarrow L}=0.00581$, with gives a rectification $R=31 \% $; (b) $T_c=0.025$, $T_h=0.325$. $J_{L\rightarrow R}=0.0499$ vs  $J_{R\rightarrow L}=0.0140$, with $R=256 \%$.}
\label{figure4}
\end{figure}

We have used again $N=20$ connected to baths of 16 thermostats each, with the same temperatures as for the homogeneous chain, and numerically solved the dynamical equations
to calculate the local temperature and the heat flux for both configurations of the baths. The interatomic potential for the regular atoms is the Morse potential (\ref{IH}).
In fig. \ref{figure4}(a), the temperature profiles show a clear asymmetry between ${L\rightarrow R}$ and ${R\rightarrow L}$. Specifically, we find $J_{L\rightarrow R}=7.6 \times 10^{-3}$ and $J_{R\rightarrow L}=5.8 \times 10^{-3}$ which gives
$R=31\%$. The effect decays with longer chains,  with $R=19\%$ for $N=100$.

\begin{figure}
\centering
\includegraphics[width=8.8cm]{{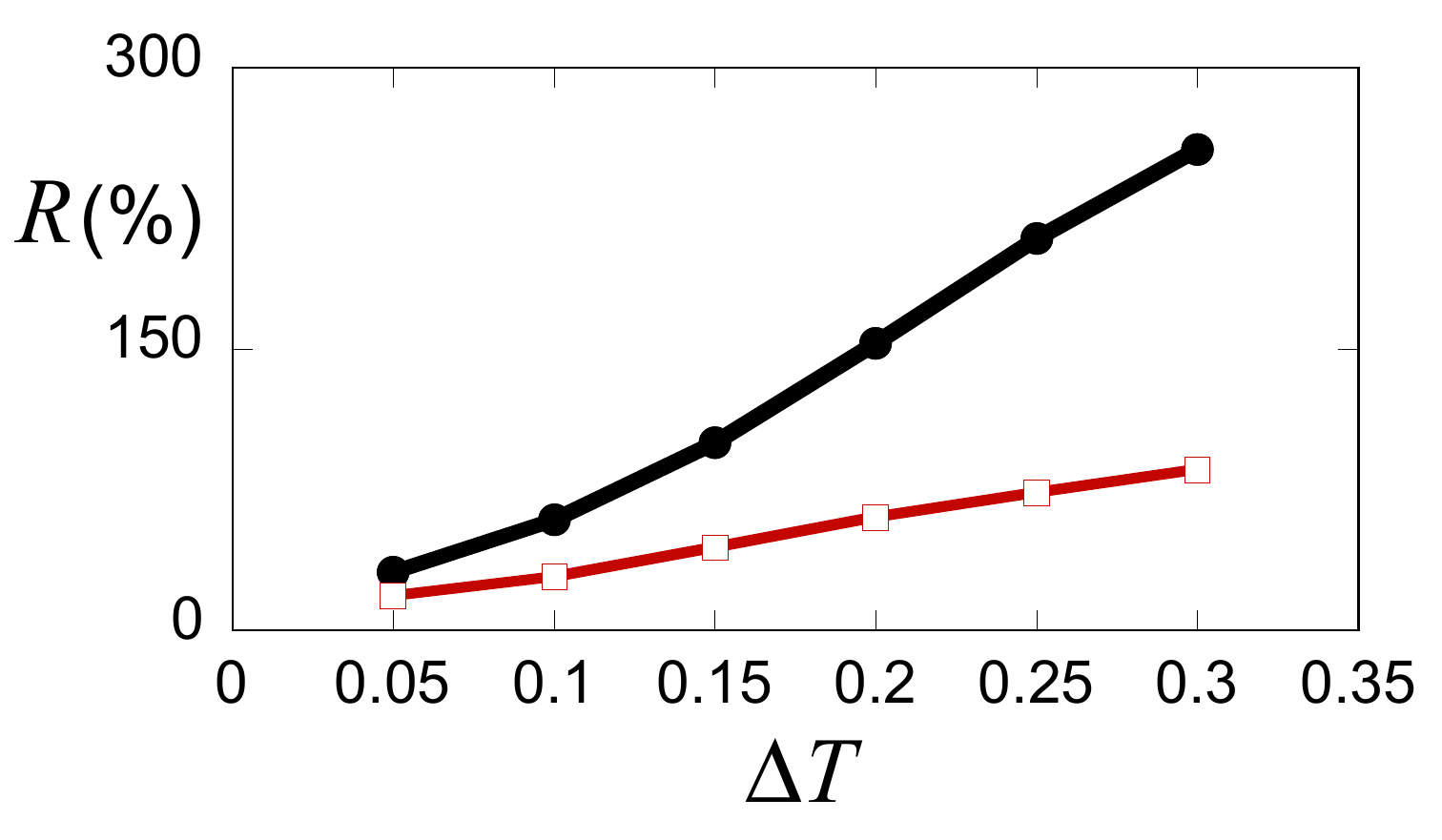}}
\caption{(Color online) Rectification factor $R$ as a function of the temperature difference between ends of the chain of atoms, $\Delta T$.
%The rectification factor shows a very strong dependency on $\Delta T$.
We have changed both $T_h$ and $T_c$ according to $T_c=0.15-(\Delta T-0.05)/2$ and $T_h=0.2+(\Delta T-0.05)/2$, with $N=20$,  keeping the rest of parameters as in fig. \ref{figure1}.
Interatomic potentials: Morse potential, eq. (\ref{IH}) (black line with circles, see the temperature profiles of extreme points in fig. 4); harmonic potential, eq. (\ref{Vhar}) (red line with squares).}
\label{figure5}
\end{figure}

This temperature profiles depend on the difference between the bath temperatures, see e.g. fig. \ref{figure4}(b). Increasing the temperature gap, but  keeping $T_h$ low enough so that the displacement of the atoms from their equilibrium positions is realistic, we find higher values of $R$. Fig. \ref {figure5} shows the strong dependence of $R$ with $\Delta T$ (black circles).
%, for the same values of the potential parameters as in fig. \ref{figure1}.
We have changed both $T_h$ and $T_c$ so that the mean temperature $(T_c+T_h)/2$ remains constant.

%Future... 100 atoms, R=0.82, worse.
%Possibility of including long range interactions? more atoms manipulated?

%%%%%%%%%%%%%%%%%%%%%%%%%%%%%%%%%%%
%%%%%%%%%%%%%%%%%%%%%%%%%%%%%%%%%%%%%%%%%

\section{Discussion}

We have presented  a scheme for thermal rectification using a one-dimensional chain of atoms which is homogeneous except
for the special interactions of one of them, the impurity, and the couplings with the baths at the boundaries.
% in such a way
%that the heat flux is different when the temperature bias of the baths at chain boundaries is reversed.
Our  proof-of-principle results for an impurity-based rectification mechanism encourage further exploration of the
impurity-based rectification, in particular
of the effect of different forms for the impurity on-site potential and its interactions with neighboring atoms.
In contrast to the majority of chain models, the structural (not due to the external bias) asymmetry in our model
is only in the impurity. The idea of a localized effect was already implicit in early works on a two-segment Frenkel-Kontorova
model \cite{Casati2004,Hu2006}, where rectification depended crucially on the interaction constant coupling between the two segments.
However, the coupling interaction was symmetrical and the asymmetry was provided by the different nature
(parameters) of the segments put in contact.
Also different from common chain models are the potentials chosen here. Instead of using the Morse
potential as an on-site model, see e.g.  \cite{Casati2002},
we have considered a natural setting where this potential characterizes the interatomic interactions,
and the on-site potential is symmetrical with respect to the equilibrium position, and actually harmonic.
The numerical results indicate that this is consistent with Fourier's law,
and also helps to isolate and identify the local-impurity mechanism for rectification.
In this regard it is useful to consider a further simplification, in the spirit of the minimalists models
proposed by Pereira \cite{Pereira2017}, so as to distill further the essence of the local rectification mechanism.
If the Morse interatomic interaction is substituted by the corresponding harmonic interaction, see fig. 1b, the rectification effect remains, albeit slightly reduced, see fig. \ref{figure5}. The chain is then perfectly linear with the only nonlinear exception  localized
again at the impurity.
The temperature dependent feature mentioned in \cite{Pereira2017} as the second necessary condition for rectification besides asymmetry, is here localized in the impurity too, and consists of a different
capability to transfer kinetic energy depending on the temperatures on both sides of the impurity.
It would be interesting to combine the impurity effect with other rectification mechanisms (such as grading, long-range interactions, or use of different segments), or with more impurities in series to enhance further the rectification effect.

Even though our motivation was to mimic the effect of a localized atom diode that lets atoms pass only one way,
unlike the atom diode \cite{Muga2004}, all interactions in the present model
are elastic. The model may be extended by adding an irreversible,  dissipative element so as to induce not only rectification but a truly Maxwell demon for heat transfer \cite{Zurek,Rus}.
On the experimental side, one dimensional chains of neutral atoms in optical lattices can be implemented with cold atoms \cite{Bloch}.
An impurity with different internal structure would be affected by a different on-site potential imprinted by a holographic mask \cite{holo}, and asymmetrical interatomic interactions
could be implemented by trapping a controllable polar molecule or mediated by atoms in parallel lattices \cite{Sabri}.

\acknowledgments

We are  indebted to G. Casati for raising our attention to thermal rectification and for providing information on his work. We acknowledge financial support by the
Basque Government (Grant No.  IT986-16) and MINECO/FEDER,UE (Grant No. FIS2015-67161-P).

%%%%%%%%%%%%%%%%%%%%%%%%%%%%%%%%%%%%%%%%%%%%%%%%%%%%%%%%%%
%                     BIBLIOGRAPHY

\bibliographystyle{eplbib}
\bibliography{Bibliograpy}

\end{document}